\documentclass[runningheads]{llncs}
\usepackage[T1]{fontenc}
\usepackage{graphicx}
\usepackage{multirow}
\usepackage{subfigure}
\usepackage{newfloat}
\usepackage{listings}
\usepackage{tabularx} 
\usepackage{multirow} 
\usepackage{array}    
\usepackage{xcolor}
\usepackage{appendix}
\usepackage{booktabs}
\usepackage{longtable}
\usepackage[utf8]{inputenc}
\usepackage{rotating}
\usepackage{afterpage}
\usepackage{placeins}
\usepackage{tabularx}
\usepackage{graphicx}

\begin{document}
\title{Echo-Teddy: Preliminary Design and Development of Large Language Model-based Social Robot for Autistic Students}
%
%
\author{
Unggi Lee\inst{1}$^\dag$ \and
Hansung Kim\inst{2} \and
Juhong Eom\inst{3} \and
Hyeonseo Jeong\inst{4} \and \\
Seungyeon Lee\inst{4} \and
Gyuri Byun\inst{5} \and
Yunseo Lee\inst{6} \and
Minji Kang\inst{7} \and
Gospel Kim\inst{8} \and \\
Jihoi Na\inst{9} \and
Jewoong Moon\inst{10}$^\dag$ \and
Hyeoncheol Kim\inst{11}$^\dag$
}
\authorrunning{Lee et al.}
%

\institute{
    Enuma, Inc., Seoul, Republic of Korea \and
    SK Telecom, Seoul, Republic of Korea \and
    Seoul Chihyeon Elementary School, Seoul, Republic of Korea \and
    Chung-Ang University, Seoul, Republic of Korea \and
    Seoul National University, Seoul, Republic of Korea \and
    University of Wisconsin-Madison, Madison, WI, USA \and
    Daegu National University of Education, Daegu, Republic of Korea \and
    Baylor University, Waco, Texas, USA \and
    Ewha Womans University, Seoul, Republic of Korea \and
    The University of Alabama, Tuscaloosa, Alabama, USA \and
    Korea University, Seoul, Republic of Korea \\
    \email{codingchild@korea.ac.kr}, \email{gnidoc327@gmail.com}, \email{juhong0422@gmail.com} \\
    \email{jeonghens@gmail.com}, \email{sarahlee0110@gmail.com}, \email{gl3013@snu.ac.kr} \\
    \email{ylee899@wisc.edu}, \email{buruburu725@gmail.com}, \email{gospel\_kim@baylor.edu} \\
    \email{bellana0309@gmail.com}, \email{jmoon19@ua.edu}, \email{harrykim@korea.ac.kr}
}

\maketitle
\begin{abstract}
Autistic students often face challenges in social interaction, which can hinder their educational and personal development. This study introduces Echo-Teddy, a Large Language Model (LLM)-based social robot designed to support autistic students in developing social and communication skills. Unlike previous chatbot-based solutions, Echo-Teddy leverages advanced LLM capabilities to provide more natural and adaptive interactions. The research addresses two key questions: (1) What are the design principles and initial prototype characteristics of an effective LLM-based social robot for autistic students? (2) What improvements can be made based on developer reflection-on-action and expert interviews?  The study employed a mixed-methods approach, combining prototype development with qualitative analysis of developer reflections and expert interviews. Key design principles identified include customizability, ethical considerations, and age-appropriate interactions. The initial prototype, built on a Raspberry Pi platform, features custom speech components and basic motor functions. Evaluation of the prototype revealed potential improvements in areas such as user interface, educational value, and practical implementation in educational settings. This research contributes to the growing field of AI-assisted special education by demonstrating the potential of LLM-based social robots in supporting autistic students. The findings provide valuable insights for future developments in accessible and effective social support tools for special education.

\end{abstract}

\section{Introduction}

Autistic students often face significant challenges in social interaction, a core aspect of their educational experience \cite{rubin2004challenges}\cite{van2015higher}. These difficulties can lead to isolation and hinder their academic and personal development \cite{whitby2009academic}. Recognizing this issue, researchers have increasingly focused on the potential of social robots as companions to help autistic students navigate social interactions \cite{lorenzo2021action}\cite{kouroupa2022use}\cite{perez2024analysis}\cite{so2018using}.

While previous studies have explored various approaches to developing social robots for autistic students, these efforts have encountered several key limitations. Many existing solutions rely on rule-based or scripted chatbots \cite{bradford2020hear}\cite{halabieh2024computer}, which lack the flexibility to adpt the unique communication needs of autistic individuals. The design of these systems often fails to consider the varied ways autistic students process and respond to social cues, leading to unnatural and ineffective interactions \cite{xygkou2024can}\cite{gu2024technological}. Additionally, the robots used in previous research tend to be expensive and difficult to mass-produce or distribute widely, limiting their accessibility and potential impact \cite{alcorn2019educators}\cite{tennyson2016accessible}. High development and production costs remain a major obstacle, restricting the scalability and widespread adoption of social robots in educational settings.

Recent advancements in artificial intelligence (AI), particularly in Large Language Models (LLMs), have created new opportunities to enhance social robot interactions. While LLMs are commonly used in education as chatbots \cite{adeshola2024opportunities}\cite{lo2023impact}, there is a growing trend toward their use in more autonomous, adaptive, and decision-making roles \cite{lee23generative}. Unlike traditional chatbots, which rely on predefined scripts and responses, LLM agents can maintain conversational context, adapt dynamically to user input, and make decisions based on real-time interactions. This agentic approach is increasingly being integrated into robotics research, enabling social robots to engage in more natural, context-aware, and personalized interactions \cite{wang2024large}\cite{kim2024survey}\cite{driess2023palm}.

To address the limitations of existing research and leverage these advancements, we introduce Echo-Teddy, an LLM agent-based social robot designed specifically for autistic students. Unlike conventional chatbot-driven robots, Echo-Teddy integrates verbal and non-verbal communication cues, allowing for more responsive and meaningful interactions. By implementing LLM agents, Echo-Teddy maintains conversational context, adapts to individual student needs, and makes nuanced decisions in real-time. Additionally, we prioritize cost-effectiveness by utilizing affordable hardware such as Raspberry Pi, ensuring that Echo-Teddy remains a scalable and accessible solution in educational robotics.

Our study focuses on two primary research questions:
\begin{itemize}
    \item RQ1: What are the design principles and initial prototype characteristics of Echo-Teddy as an LLM agent-based social robot?
    \item RQ2: How can the initial prototype of Echo-Teddy be improved based on developer reflection-on-action and expert feedback?
\end{itemize}

\subsection{Contribution}

By exploring these questions, our research contributes to the development of more accessible, effective, and scalable social support tools for autistic students. Echo-Teddy not only addresses immediate limitations in social robot design but also lays the groundwork for future innovations at the intersection of artificial intelligence, robotics, and special education. In particular, our iterative evaluation process—combining developer insights and expert feedback—ensures that future versions of Echo-Teddy are better aligned with the needs of autistic students and the realities of special education environments.

The key contributions of this research are:
\begin{itemize}
    \item Development of Echo-Teddy, a novel LLM agent-based social robot specifically designed to address the limitations of previous chatbot-based solutions for autistic students.
    \item Exploration of LLM agent capabilities in educational robotics, demonstrating their potential for more adaptive, context-aware, and personalized interactions.
    \item Identification of key design principles for effective and scalable LLM agent-based social robots in special education.
    \item Implementation of a cost-effective prototype, leveraging affordable hardware like Raspberry Pi, to demonstrate the feasibility of accessible and scalable robotics solutions.
\end{itemize}

\begin{figure*}[hbt!]
    \centering
    \includegraphics[width=1\textwidth]{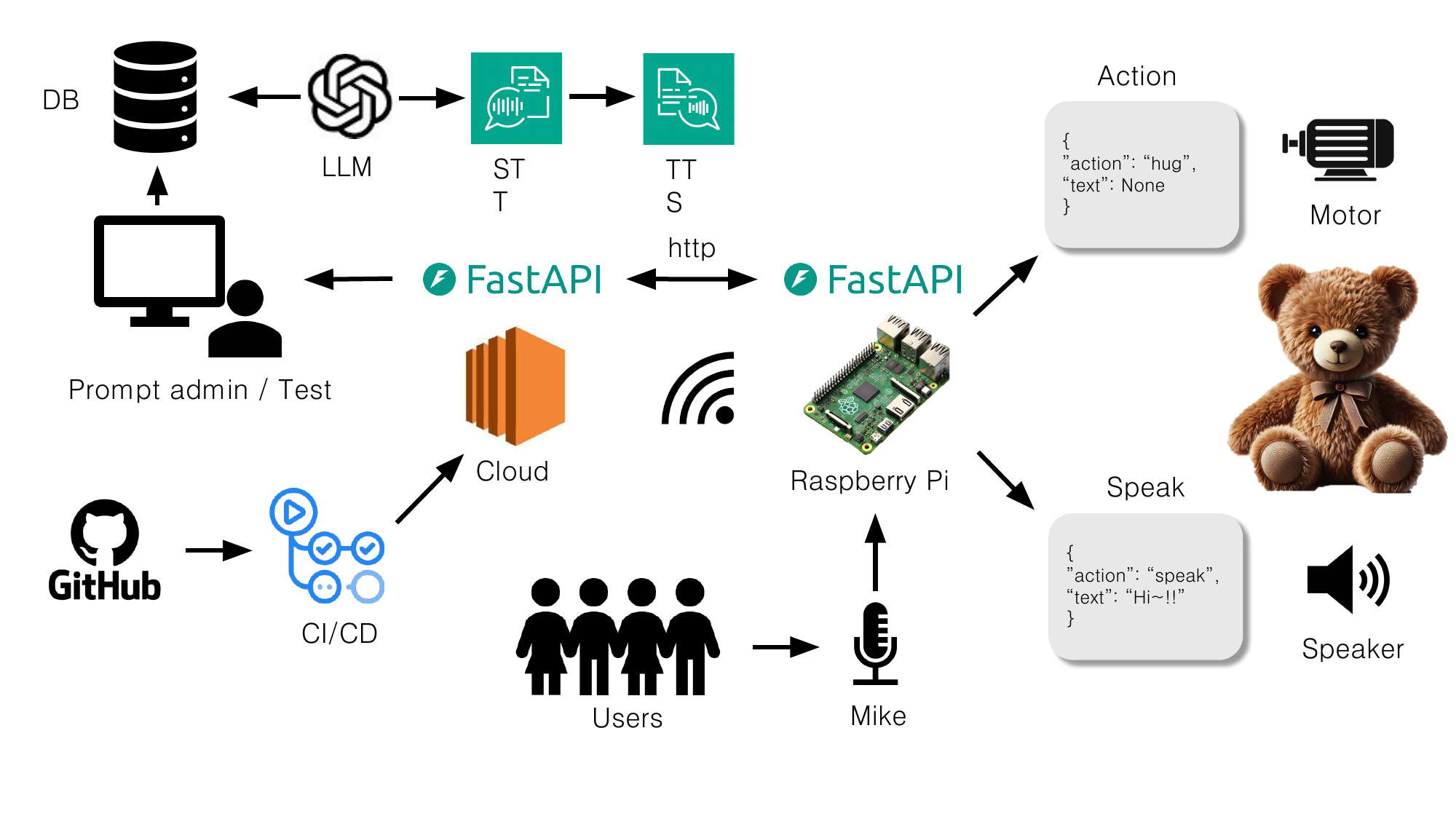}
    \caption{System architecture of echo teddy.}
    \label{fig:echo-teddy-arch}
\end{figure*}

\section{Related Work}

\subsection{Social Robotics for Autistic Students} 

The field of robotics has been widely explored to enhance skill development in autistic children. These technologies are primarily used to support learning and social development, with a particular focus on improving and assessing social communication skills. A variety of robotic forms, including both physical and virtual robots, have been utilized for this purpose \cite{Mohammed2020robot}.

Socially Assistive Robotics (SAR) represents a rapidly growing research field that focuses on designing and implementing robots that assist users through social interaction rather than physical engagement \cite{Diehl2012cu}\cite{Ricks2010tc}\cite{Syriopoulou-Delli2020ra}. SARs employed in autism intervention programs vary widely in form and function, including humanoid, animal-like, and machine-like systems, each differing in physical design and realism. Notable SARs include KASPAR, ZENO, Probo, and ZECA, which have been used in multiple studies \cite{Syriopoulou-Delli2020ra}. Additionally, commercial AI-powered robots, such as Nao and Pepper, have been introduced as communicative facilitators to assist autistic children in social interactions \cite{10.1145/3461615.3485421}\cite{Lemaignan2024pepper}. However, due to the limited availability of commercially suitable robotic platforms for autism intervention, many research groups have opted to develop custom-designed robots tailored to specific intervention needs. 

Studies have demonstrated positive effects of SARs on social behaviors in autistic children, including increased eye contact, enhanced verbal communication, improved imitation skills, greater physical engagement, enriched play interactions, reduced stereotypical behaviors, and heightened emotional responses \cite{Scassellati2012ar}.

Despite the promising potential of Socially Assistive Robotics (SARs) for autistic children, only a limited number of studies have investigated their long-term, continuous application \cite{conti2017dd}\cite{vagnetti2024social}. Most research has focused on short-term interventions, leaving gaps in understanding their sustained effectiveness in real-world environments. Therefore, ongoing evaluation of their reliability, adaptability, and usability in daily educational and therapeutic settings is essential. Further studies are needed to examine SAR effectiveness across diverse demographic factors, including sex, age, and cognitive ability \cite{Mohammed2020robot}. In addition, to ensure broad accessibility and widespread adoption, SARs must be affordable, scalable, and designed for everyday use in practical learning environments with daily settings. 

\subsection{Large Language Model for Robotics}

Recent advancements in robotics have increasingly leveraged Large Language Models (LLMs) to enhance robotic capabilities across various domains. LLMs such as GPT-4 and similar models are now being integrated into key robotic functions, including communication, perception, planning, and control \cite{wang2024large}\cite{kim2024survey}\cite{zeng2023large}. These integrations aim to equip robots with advanced cognitive abilities, enabling them to understand, process, and execute complex tasks based on natural language instructions.

Several notable examples highlight the transformative role of LLMs in robotics. RT-1, SayCan, and PaLM-E have demonstrated how robots can perform manipulation tasks, such as picking up, placing, and moving objects based on human instructions. The RT series (RT-1, RT-2, RT-X) has shown that robots can learn from large-scale data to generalize across a wide range of tasks \cite{DBLP:conf/rss/BrohanBCCDFGHHH23}\cite{pmlr-v229-zitkovich23a}\cite{vuong2023open}. SayCan integrates LLMs with robotic affordances, enabling context-aware task planning and execution in real-world environments \cite{Ahn2022DoAI}. PaLM-E further extends this by incorporating visual and perceptual inputs, allowing robots to reason and act based on both textual and environmental cues \cite{Driess2023PaLMEAE}.

Beyond research prototypes, companies like Figure AI and Tesla have introduced advanced humanoid robots, such as Figure AI’s autonomous humanoid and Tesla’s Optimus Gen 2, which demonstrate high-performance robotic interactions that may be enhanced by LLMs \cite{figure_ai}\cite{tesla_optimus_gen2}. These developments reflect a growing trend toward integrating LLMs into robotic systems, allowing for more natural, interactive, and intelligent behaviors.

In educational settings, studies have explored the use of LLMs to teach robotics-related content, enhancing learning experiences through interactive dialogues, personalized tutoring, and real-time explanations \cite{Kahl2024}\cite{shu2024llms}. However, research on integrating LLMs directly into educational robotics remains limited. The potential for LLMs to personalize learning, simulate complex robotic behaviors, and facilitate real-time problem-solving in robotics education remains a largely unexplored but promising area for future research.

\section{Methodology}

\subsection{Research Procedure}

This study employed a systematic design and evaluation process to develop Echo-Teddy, an LLM agent-based social robot for autistic students. The research proceeded in three phases: reviewing existing studies, developing the prototype, and conducting expert evaluations to refine its design and functionality. The first phase involved a comprehensive review of prior research on social interaction support for autistic students and robot-assisted interventions. This review informed the key design principles guiding the development of Echo-Teddy. The design emphasized context-aware conversational adaptability using an LLM, voice-based interaction for natural communication, non-verbal cues to enhance engagement, and a cost-effective hardware structure to ensure scalability.
Second, based on these principles, a prototype was developed using Raspberry Pi as the hardware platform, integrating LLM-based conversational software to facilitate interactive dialogue. The system was designed to process voice input and generate responses using synthesized speech. Additionally, simple non-verbal gestures were incorporated to support engagement, allowing the robot to communicate using both verbal and physical cues. Last, following the prototype development, reflective evaluation by the development team and expert interviews were conducted to assess Echo-Teddy’s usability and identify areas for improvement. Five special education experts participated in semi-structured interviews, evaluating aspects such as response speed, physical design, non-verbal communication, and educational applicability. The qualitative analysis of these evaluations identified key areas for refinement, informing subsequent improvements to the prototype. Insights from these findings shaped the next iteration of Echo-Teddy, ensuring better alignment with the needs of autistic students and practical classroom implementation.

\subsection{Research Tools}

\subsubsection{Hardware and Software}

Echo-Teddy integrates hardware and software components to create an interactive system designed specifically for autistic students. The hardware is built around a Raspberry Pi 5, which serves as the central processing unit. It is connected to a microphone for voice input, a speaker for audio output, and a motor for basic physical gestures. These components are enclosed within a child-friendly, soft teddy bear exterior, ensuring both a comforting design and practical durability. The use of affordable and widely available hardware components enhances cost-effectiveness and scalability, making Echo-Teddy a viable solution for broader implementation. Additional hardware specifications are detailed in Appendix 1.

The software architecture integrates cloud-based AI services to enhance interaction. Amazon Web Services (AWS) processes Speech-to-Text (STT), converting verbal input into text for further analysis. OpenAI’s GPT-4o-mini generates responses dynamically, allowing for context-aware and adaptive conversations that align with the user’s input and behavioral patterns. To accommodate Korean-speaking users, Naver Clova Voice is used for Text-to-Speech (TTS), ensuring natural and engaging auditory output. A prompt management module structures the interaction, refining the model’s responses to align with social appropriateness and individual student needs. The backend, built using FastAPI and hosted on AWS, facilitates efficient communication between the hardware and cloud-based AI services, ensuring scalability, low-latency processing, and reliable performance in real-time interactions.

\subsubsection{Interview Questionnaire}

The interview questionnaire for Echo-Teddy evaluation were designed to cover four key categories: \textit{Affordance}, \textit{Usability}, \textit{Instructional Design}, and \textit{Instructional Usefulness}. These categories were chosen to provide a comprehensive assessment of the tool from various perspectives. The \textit{Affordance} focuses on initial perceptions and understanding of the tool's functions. \textit{Usability} address the overall satisfaction, functionality, design, and interactive aspects such as sound and movement. The \textit{Instructional Design} delves into the educational context, exploring potential use scenarios, learning processes, and expected outcomes in terms of behavior, knowledge, skills, and attitudes. Finally, the \textit{Instructional Usefulness} examines the tool's effectiveness in achieving learning goals, its appropriateness for different learners, and potential benefits, both intended and unexpected. This structured approach to the interview questions ensures a thorough evaluation of Echo-Teddy, covering both its technical aspects and its educational value. By addressing these diverse areas, the interviews aim to gather comprehensive insights that will inform future improvements and adaptations of the tool, ultimately enhancing its effectiveness in supporting autistic students.

\begin{table*}[hbt!]
\centering
\scriptsize
\begin{tabular}{p{3cm}p{3cm}p{6cm}}
\hline
\multicolumn{1}{c}{\textbf{Category}} & \multicolumn{1}{c}{\textbf{Aspect}} & \multicolumn{1}{c}{\textbf{Question}} \\
\hline
\multirow{2}{*}{Affordance} & Use intention & What did you think you would want to do when you first saw the tool? \\
\cline{2-3}
 & Function understanding & Explicit vs. implicit function understanding? \\
\hline
\multirow{5}{*}{Usability} & Satisfaction & Overall satisfaction with the tool? \\
\cline{2-3}
 & Functionality & Is the functionality appropriate? \\
\cline{2-3}
 & Design & Is the visual design appropriate? \\
\cline{2-3}
 & Sound/Voice & Is the sound/voice of the tool appropriate? \\
\cline{2-3}
 & Movement & Is the movement of the tool appropriate? \\
\hline
\multirow{8}{*}{Instructional Design} & Use context & Where can the tool be used? \\
\cline{2-3}
 & Alternative methods & Other ways to achieve the learning goal? \\
\cline{2-3}
 & Learning process & What is the ideal learning process with the tool? \\
\cline{2-3}
 & Behavior outcome & What behavior change should occur? \\
\cline{2-3}
 & Knowledge outcome & What key knowledge should be learned? \\
\cline{2-3}
 & Skill outcome & What skills should be learned? \\
\cline{2-3}
 & Affective outcome & What attitudes or thoughts should change? \\
\cline{2-3}
 & Information format & How should the tool present information? \\
\hline
\multirow{11}{*}{Instructional Usefulness} & Learning process & What should the learning process look like? \\
\cline{2-3}
 & Behavior outcome & What behavior change will occur? \\
\cline{2-3}
 & Knowledge outcome & What knowledge will be learned? \\
\cline{2-3}
 & Skill outcome & What skills will be learned? \\
\cline{2-3}
 & Affective outcome & What attitudes or thoughts will change? \\
\cline{2-3}
 & Validity & Is the tool effective in achieving goals? \\
\cline{2-3}
 & Generality & Is the tool appropriate for all learners? \\
\cline{2-3}
 & \multirow{3}{*}{Usefulness} & Does the tool have any benefits? \\
 & & - What intended benefits are there? \\
 & & - What unexpected benefits are there? \\
\cline{2-3}
 & Improvements & What needs improvement and why? \\
\hline
\end{tabular}
\caption{Interview Questions for Echo-Teddy Evaluation}
\end{table*}

\section{Result}

\subsection{RQ1: What are the design principles and initial prototype characteristics of Echo-Teddy?}

\subsubsection{Design principles of Echo-Teddy}

The design principles of Echo-Teddy are structured to comprehensively support social and social emotional development in autistic children, focusing on four key themes: Potential User, Ethical Consideration, Customization, and Usage. The robot is designed to function as both a peer and an assistant, accommodating the neurodiverse characteristics of autistic children, including gaze aversion, preference for structured patterns, heightened perceptual sensitivity, and attention to hierarchical information and fine details. By integrating these elements, Echo-Teddy creates an interactive, supportive, and engaging experience tailored to the needs of its users.

Echo-Teddy’s verbal and behavioral output aligns with best practices in autism support, ensuring that its speech patterns are age-appropriate and incorporate evidence-based teaching strategies. These strategies reinforce positive behaviors while mitigating interfering behaviors. The robot is designed to elicit and encourage target verbal and behavioral responses, offering structured interactions while maintaining the flexibility needed to adapt to individual learning preferences. Unlike humanoid robots, Echo-Teddy features a soft, fur-covered exterior, a design choice aimed at reducing anxiety and increasing comfort for autistic children who may find highly realistic or human-like robotic faces overwhelming.

Ethical considerations are central to Echo-Teddy’s development. The robot consistently uses positive reinforcement strategies to prevent frustration and enhance engagement. Its speech and behavioral responses dynamically adjust to accommodate each child's individual needs and comfort levels. A key objective in its design is to minimize stressors while fostering an environment that encourages meaningful interaction.

Customization plays a pivotal role in Echo-Teddy’s adaptability. Caregivers and educators can tailor its communication style, interaction topics, and behavioral prompts to align with each child’s unique preferences and developmental goals. This customization extends beyond software, as the robot’s physical appearance can be modified to better suit individual sensory and aesthetic preferences.

Practical usability is another core aspect of Echo-Teddy’s design. The robot is built to be durable and resilient, capable of withstanding minor impacts or water exposure, ensuring longevity in varied educational and home environments. Additionally, Echo-Teddy is designed for independent operation, eliminating the need for constant human intervention and allowing children to interact with it autonomously.

By integrating neurodiverse-friendly interaction models, ethical safeguards, extensive customization options, and practical durability, Echo-Teddy is designed to be an effective and accessible tool for enhancing social communication skills in autistic children. These principles ensure that Echo-Teddy is not only tailored to the unique needs of its users but also remains ethically responsible, adaptable, and functional in real-world applications.

\begin{table*}[hbt!]
\scriptsize
\begin{tabular}{p{2cm} p{3cm} p{7cm}}
\hline
\multicolumn{1}{c}{\textbf{Themes}} & \multicolumn{1}{c}{\textbf{Categories}}   & \multicolumn{1}{c}{\textbf{Design Principles}}                                                                                                   \\
\hline
Potential User                      & Purpose of the Robot                      & This robot should mainly improve the social/socio-emotional skills of autistic children by performing social communication and interaction. \\
                                    &                                           & The robot should act like a facilitator, including peers and assistants.                                                                         \\
                                    & Characteristics of autistic students      & The robot should be designed considering the neurodiverse characteristics of autistic children. (gaze aversion, pattern recognition, perceptual processing, and exceptional focus for paticular topic).
                                    \\
                                    & Output of the robot (Verbal)              & Keep the utterance style and length appropriate to a child of the same age as the user.                                                          \\
                                    &                                           & The robot should activate verbal teaching strategies to induce positive behavior and reduce interfering behaviors.                               \\
                                    &                                           & The robots should be able to elicit the target verbal behavior in children with autism.                                                             \\
                                    & Output of the robot (Behavioral)          & The robot should activate behavioral teaching strategies to induce positive behavior and reduce interfering behaviors.                           \\
                                    &                                           & The robots should be able to elicit the target behavior in children with autism.                                                                    \\
                                    & Appearance of the robot                   & The appearance of the robot should be non-humanoid.                                                                                              \\
\hline
Ethical Consideration               &                                           & Avoid frustration by using positive feedback.                                                                                                    \\
                                    &                                           & Reflect the unique needs of the user in your speech and actions.                                                                                 \\
                                    &                                           & Minimize stress sources and make the participants comfortable.                                                                                   \\
\hline
Customization                      & Preference of the robot                   & The preference of the robot should be customized by the caregiver or instructor of the user.                                                     \\
                                    & Output of the robot (Verbal)              & The subject of the communication should be customized.                                                                                           \\
                                    & Output of the robot (Behavioral)          & The set of behaviors to be stimulated in the child should be customized by the caregiver or the instructor.                                       \\
                                    & Appearance of the robot                   & The appearance of the robot should be customizable.                                                                                              \\
\hline
Usage                               &                                           & Consider the context of use to ensure appropriate sturdiness and prevent damage from shocks or water; the user environment must be considered.   \\
                                    &                                           & The robot should not need additional human support during use.\\
\hline
\end{tabular}
\caption{Design principle for Echo-Teddy.}
\end{table*}

\subsubsection{Initial prototype of Echo-Teddy}

\begin{figure*}[hbt!]
    \centering
    \includegraphics[width=1\textwidth]{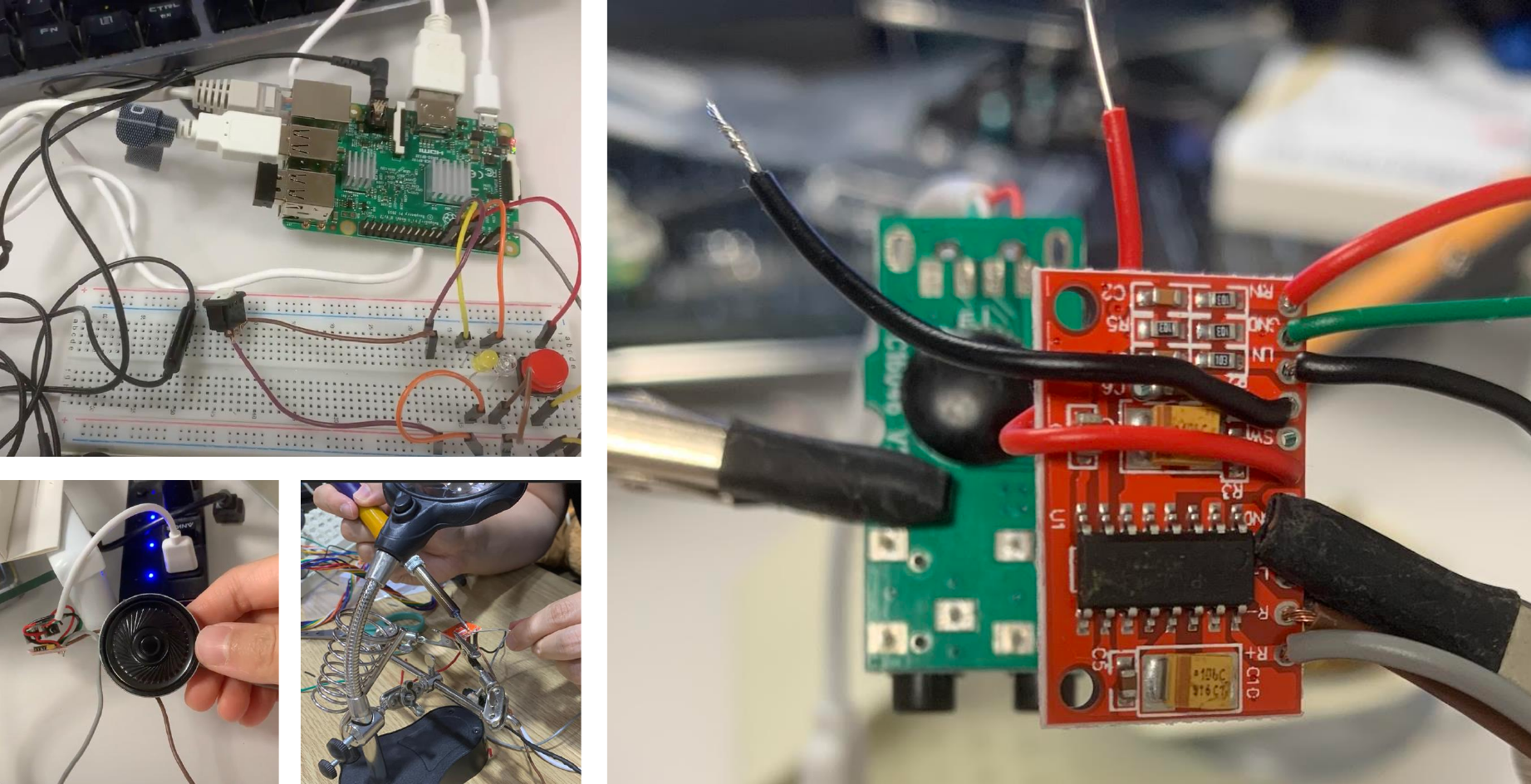}
    \caption{To reduce costs, we purchased microphone and speaker components separately and assembled them using soldering techniques.}
    \label{fig:echo-teddy-components}
\end{figure*}

The hardware design of Echo-Teddy is built on a Raspberry Pi platform, chosen for its cost-effectiveness, scalability, and ability to support real-time interaction. The system efficiently transmits audio files and action commands between the Raspberry Pi and the server, ensuring low-latency communication for natural conversations. To further reduce production costs, the microphone and speaker were assembled using custom-purchased components and soldering techniques (Figure \ref{fig:echo-teddy-components}). This approach allowed for greater flexibility in hardware integration while maintaining affordability for broader implementation.

\begin{figure*}[hbt!]
    \centering
    \includegraphics[width=1\textwidth]{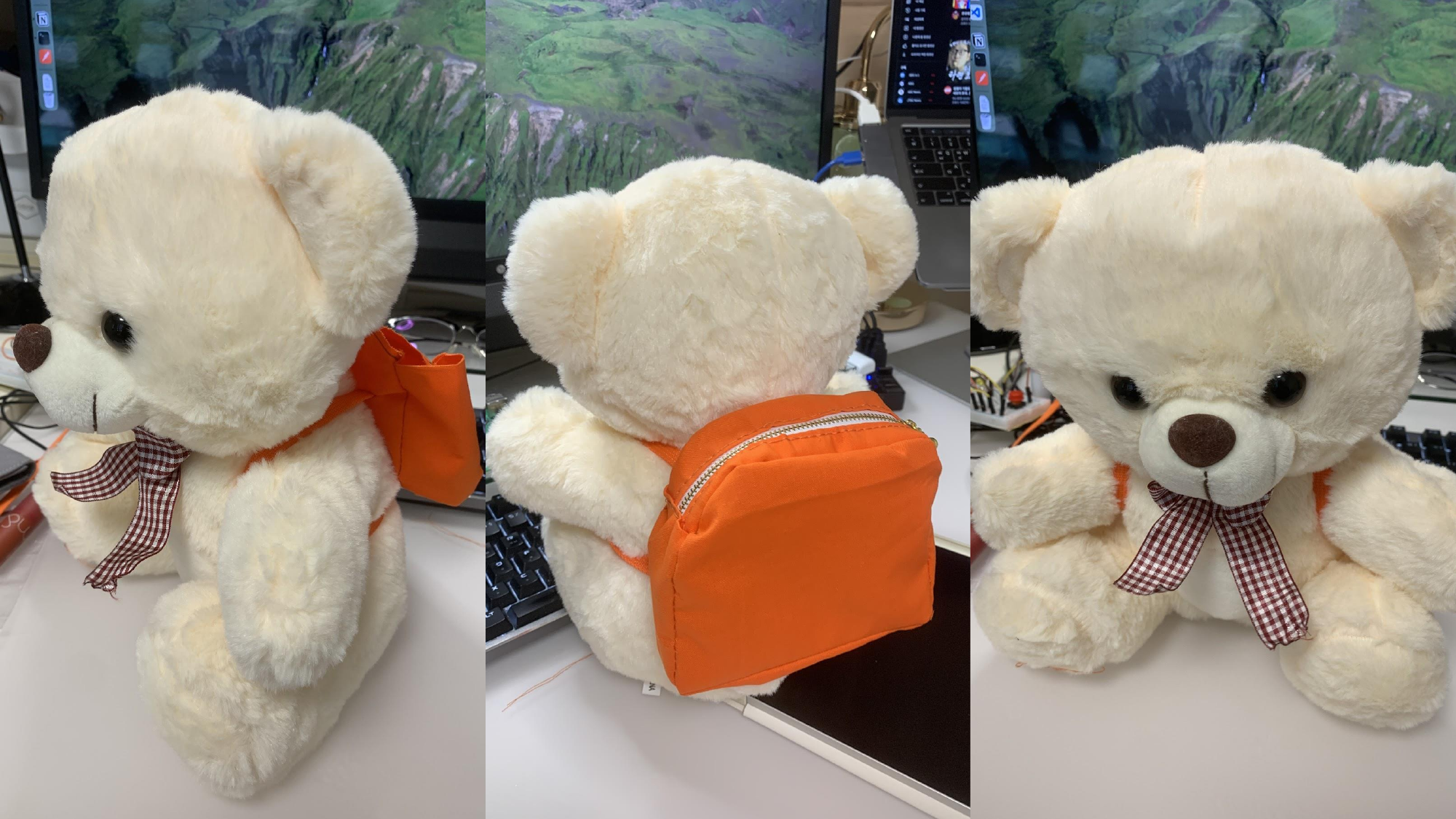}
    \caption{To prevent the heat buildup, we made backpack to contain the Raspberry Pi.}
    \label{fig:echo-teddy-backpack}
\end{figure*}

During the production process, it was observed that placing the Raspberry Pi inside the plush doll led to heat buildup, which caused performance degradation. To address this, a backpack-style enclosure was designed to house the Raspberry Pi externally, allowing for better heat dissipation without compromising portability (Figure \ref{fig:echo-teddy-backpack}). This design also improves ease of maintenance and accessibility for future hardware upgrades.

\begin{figure*}[hbt!]
    \centering
    \includegraphics[width=0.5\textwidth]{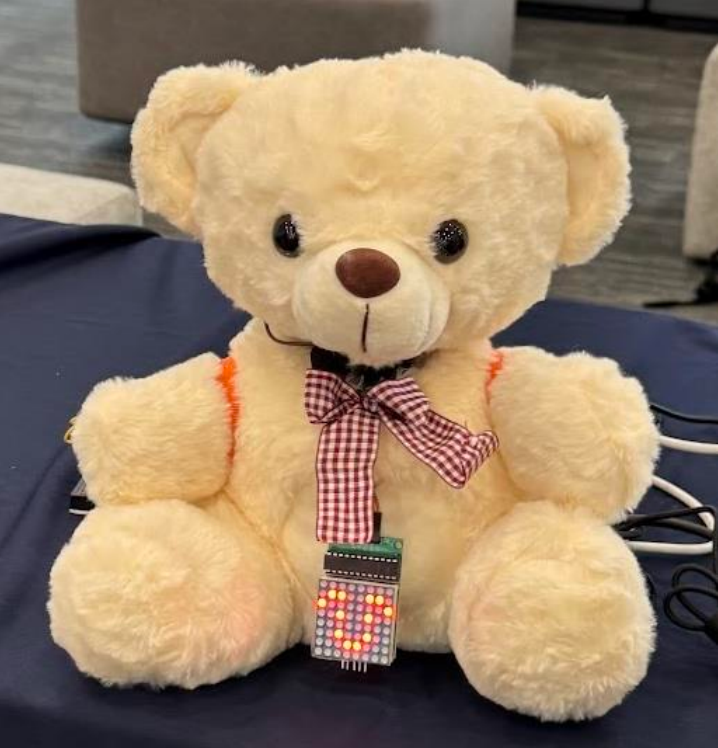}
    \caption{We used dot matrix to express the emotions of Echo-Teddy.}
    \label{fig:echo-teddy-dot-matrix}
\end{figure*}

The initial prototype version integrates attached motors to enable basic movements, such as nodding, providing simple nonverbal communication cues. Currently, the range of motion is limited to head movements and facial expressions, which are displayed using a dot matrix (Figure \ref{fig:echo-teddy-dot-matrix}). These features are intended to enhance emotional expressiveness and engagement in interactions with users. 

For connectivity, the system utilizes the built-in Wi-Fi module of the Raspberry Pi, ensuring stable access to cloud-based services. Additionally, mobile phone tethering is available as an alternative network solution, enabling portability across different settings, including classrooms, therapy environments, and home use. This flexible connectivity setup ensures that Echo-Teddy remains accessible and functional in diverse user environments.

The server system of Echo-Teddy integrates advanced cloud-based technologies to ensure efficient, scalable, and personalized interactions for autistic students. At its core, the system relies on OpenAI’s API for dialogue generation and natural language processing, allowing for context-aware and adaptive conversational interactions. To enhance flexibility and control over language model interactions, a prompt management module has been implemented, enabling LLM administrators to easily update and manage prompt texts for fine-tuned responses.

For speech processing, Echo-Teddy employs AWS Transcribe for Speech-to-Text (STT) functionality, ensuring accurate and efficient transcription of user input. Text-to-Speech (TTS) is handled by both AWS Polly and Naver Clova Voice, with Naver Clova Voice specifically incorporated to support Korean-speaking users. This multi-engine approach ensures high-quality, natural speech synthesis across different languages and user preferences.

The server infrastructure is built using FastAPI, following RESTful API principles, and is hosted on AWS EC2 to maintain reliable and scalable operations. To streamline development and deployment, the system integrates a CI/CD pipeline using GitHub Actions, enabling automated testing, integration, and deployment, reducing maintenance overhead and improving system stability. A key feature of the backend is its ability to transmit audio files along with structured action commands in JSON format, allowing for synchronized multimodal interactions. This ensures that Echo-Teddy’s speech output is coordinated with its physical gestures, enhancing engagement and communication effectiveness.

By leveraging this comprehensive cloud-based system architecture, Echo-Teddy provides a stable, adaptive, and scalable interaction platform designed to support the diverse communication needs of autistic students. The integration of modular and configurable components further ensures that the system remains flexible and customizable, meeting the evolving demands of special education applications.

\subsection{RQ2: What improvements can be made to the initial prototype of Echo-Teddy based on developer reflection-on-action and interviews with experts?}

\subsubsection{Reflection-on-action of developers}

This section reflects on the design and implementation decisions made during the development of the initial prototype of Echo-Teddy, highlighting key challenges encountered and the provisional solutions proposed. These insights provide a foundation for future refinements, ensuring that the robot effectively supports social and emotional development in autistic students.

One of the primary areas of focus was enhancing nonverbal communication capabilities to improve the naturalness of interaction. While the initial prototype allowed Echo-Teddy to express emotions based on user input, its expressive range was limited to seven predefined facial expressions. To increase emotional depth and adaptability, an additional feature involving movable eyebrows could be introduced, allowing for more nuanced and dynamic facial expressions. This modification would enable the robot to better reflect emotional subtleties, making interactions more engaging and responsive to diverse social cues.

A second area of improvement involved implementing a monitoring system for caregivers and educators. Since interactions with Echo-Teddy can provide valuable insights into a child's social communication patterns, enabling convenient tracking of interaction history is essential for individualized intervention planning. To facilitate this, experts recommended optimizing the website for mobile accessibility or developing a dedicated mobile application, allowing caregivers and educators to easily review and analyze communication logs. 

Third, ensuring the reliability of Echo-Teddy’s emotional responses was another critical challenge. Accurate and contextually appropriate emotional reactions to user input are essential for fostering both self-awareness and emotional comprehension in autistic students. However, facial expressions generated from new data should undergo rigorous evaluation to verify accuracy and appropriateness, ensuring that Echo-Teddy's emotional feedback is both credible and developmentally supportive. As noted by Rawal et al. \cite{rawal2022facialemotionexpressionshumanrobot}, effective human-robot interactions depend on accurate emotion recognition and expression, reinforcing the need for continuous validation and refinement of Echo-Teddy’s emotional modeling.

Another key consideration was adapting Echo-Teddy’s appearance to strengthen emotional bonds with users. The physical design of social robots plays a significant role in user engagement and comfort. Research by Ricks and Colton \cite{5509327} suggests that while autistic children benefit from both humanoid and non-humanoid robots, they tend to engage more actively with non-humanoid designs. This finding supports the idea that flexibility in physical form—such as allowing for different shapes or textures—may improve user preference and interaction quality. Providing customizable external features, such as different animal forms or textures, could further optimize engagement based on individual sensory preferences.

Reflecting on these early design decisions has yielded valuable insights into both project management and technical execution. Future development efforts will focus on enhancing Echo-Teddy’s real-time emotional expressiveness, refining its monitoring capabilities, validating emotion-driven responses, and expanding customization options. These refinements will ensure that Echo-Teddy remains an adaptive, engaging, and effective tool for supporting social and emotional development in autistic students.

\subsubsection{Results of interview with experts}

The interview included experts with extensive experience in special education who provided diverse perspectives on the early prototype of the LLM-based social robot Echo-Teddy for students with autism spectrum disorder. The content of the interview was primarily analyzed in terms of seven themes: (1) response speed (processing delay), (2) the robot’s physical form and external features, (3) interaction goals and usage scenarios, (4) nonverbal communication and modeling strategies, (5) the range and characteristics of the target students, (6) generalizability and additional considerations, and (7) implementation costs and scalability. Below is a summary of the experts’ key opinions and improvement suggestions for each theme.

First, experts pointed out that Echo-Teddy experienced a delay of approximately 6–10 seconds when communicating with external servers for STT and TTS. They emphasized that immediate feedback is crucial for students with autism, as delays can lead to loss of attention or increased anxiety during communication. They noted that communication skills must be reinforced through repeated, prompt interactions. To address this, they suggested either (1) upgrading hardware to support faster local processing (e.g., Jetson Nano) or (2) splitting audio data into smaller segments and synthesizing speech sequentially for playback.

Second, the interview included discussions about the rationale behind Echo-Teddy being designed as a teddy bear and the appropriateness of this choice. Experts acknowledged that "soft plush robots can be a positive approach for certain students, as safety and sensory preferences are critical for students with autism." However, they also recommended offering modular options, such as different animal shapes (e.g., dinosaurs), to accommodate students who may react negatively to certain textures or appearances. Furthermore, some experts highlighted the importance of clear visual focus, such as the size and positioning of the robot’s eyes, to encourage eye contact. However, others cautioned against prioritizing eye contact as a universal goal, as it may provoke anxiety or discomfort for some students. Ethical considerations and individual differences were emphasized.

Third, while the development team initially focused on scenarios where students would engage in one-on-one conversations with the robot under the supervision of caregivers or teachers, experts emphasized the importance of fostering peer interactions. They suggested that Echo-Teddy could serve as a mediator to encourage participation and extend social skills in group settings, such as inclusive classrooms. They proposed that the robot could facilitate dialogue between autistic students and their peers, promoting social engagement. However, they also noted potential challenges, such as peers perceiving the robot as a novelty or the risk of students overly depending on the robot.

Fourth, experts emphasized the importance of modeling nonverbal communication (e.g., nodding, gestures, gaze direction) for students with autism, as these skills are as critical as verbal language. They suggested that Echo-Teddy could perform human-like gestures, such as turning its head, waving, or nodding, to encourage imitation behaviors. However, they advised against certain actions, like "hugging," which may not align with typical daily interactions or could cause sensory discomfort. Additionally, they discussed whether immediate reinforcement (e.g., the robot saying, "Good job!") should be provided after modeling or if maintaining a natural conversational flow would be more effective. Designing consistent teaching strategies and robot responses was deemed essential.

Fifth, the experts recommended focusing initially on "high-functioning autistic students who can use spoken language" while also exploring the feasibility of extending Echo-Teddy to students using AAC (augmentative and alternative communication) devices. Some students rely on picture icons or switch operations to express themselves instead of direct speech. Therefore, they emphasized that the robot should not be limited to processing verbal speech alone. If the robot could recognize and respond to electronic voices generated by AAC devices, it could benefit a broader spectrum of students.

Sixth, the experts stressed the importance of ensuring that interactions with the robot could generalize to real-life conversations with other humans. They cautioned that learning communication skills with the robot might have limited impact if not transferred to interactions with peers or family members. Skills like eye contact or body orientation were highlighted as "common competencies" that should be modeled by the robot and reinforced by caregivers or teachers. Practical considerations, such as the robot’s durability, hygiene, and connectivity, were also mentioned as essential factors for long-term use.

Finally, experts noted the financial constraints faced by special education programs, where the budget for equipment is often limited. They suggested that Echo-Teddy could be developed as an open-source project, allowing institutions to build DIY versions tailored to their specific needs. This approach could reduce costs and enable schools or organizations to customize the robot’s appearance, voice, and features to meet diverse sensory and preference requirements.

\section{Discussion}

The expert interviews provided key insights into refining Echo-Teddy’s design and functionality to better support social communication in autistic students. Experts highlighted the need for immediate responsiveness, a carefully designed physical and behavioral model, goal-oriented interaction strategies, and practical implementation considerations. Their feedback emphasized the importance of ensuring that Echo-Teddy not only facilitates meaningful interaction but also generalizes its impact to real-world social settings.

First, one major finding was the critical importance of real-time responsiveness in maintaining engagement for autistic students. Experts observed a 6–10 second processing delay due to cloud-based speech recognition and text-to-speech generation, which they noted could increase anxiety and disrupt interaction flow. Given that immediate and predictable feedback is essential for reinforcing communication skills, delays risk causing students to lose focus or disengage from the conversation. Cano et al. \cite{cano2023design} found that latency reductions in robotic responses not only enhance engagement but also mitigate behavioral disruptions, making this a critical area for refinement. Implementing edge computing strategies or optimizing speech processing pipelines may be essential for achieving the low-latency performance necessary to support effective and anxiety-free interactions for autistic students.

Second, the robot’s form and behavioral modules must balance the individual characteristics of autistic students with educational goals, particularly in distinguishing between modeling social behaviors and facilitating natural conversation. In inclusive classroom settings, incorporating subtle nonverbal cues such as head nods, body tilts, and gaze direction can encourage peer interaction by reinforcing appropriate social responses. However, ensuring design flexibility is crucial, as autistic students have diverse sensory preferences and interaction styles. Tailoring the robot’s activities, gestures, and verbal feedback based on user preferences and real-time adaptation is particularly important. Lee and Park  \cite{maroto2024personalizing} demonstrate that personalization enhances both engagement and intervention effectiveness, reinforcing the importance of customizable interaction strategies.Similarly, Cano et al. \cite{cano2023design} emphasize that user-centered design ensures robots remain effective across a spectrum of student needs, cautioning against one-size-fits-all approaches. For instance, while verbal reinforcement strategies (e.g., the robot saying "Good job!") can be effective for some students, others may find explicit praise stressful or disruptive. To prevent additional stress or confusion, reinforcement should be context-aware, ensuring that it aligns naturally with the conversation flow rather than feeling overly scripted or intrusive. This highlights the need for adaptive response mechanisms, allowing the robot to modulate its speech and gestures based on the child's individual comfort levels and engagement patterns.

Third, to broaden the scope of Echo-Teddy, ensuring scalability and generalizability is essential. Skills learned through robot-assisted interactions should seamlessly transfer to real-life settings, enabling autistic students to apply these communication skills with peers, caregivers, and teachers. Without this transferability, the benefits of robot-mediated interactions risk being isolated to controlled environments rather than supporting meaningful social engagement in daily life.  Choi et al \cite{santos2023applications} emphasize that scalable robot designs enable broader applicability across diverse educational and social settings, ensuring that interventions are not limited to small-scale experiments. A structured learning approach—beginning with caregiver-led activities and progressively expanding to peer interactions—has been shown to enhance social skill transfer beyond interactions with the robot itself. Clabaugh et al. \cite{clabaugh2019long} further highlight the value of long-term personalized interventions, demonstrating that sustained engagement reinforces skill retention and generalization. Such findings indicate that for Echo-Teddy to achieve wide-scale adoption, it must integrate Augmentative and Alternative Communication (AAC) devices, ensuring compatibility with students who rely on visual icons, text-based communication, or switch-access systems. Furthermore, the robot must be equipped with robust network capabilities to ensure seamless connectivity in classroom and home environments. Durable hardware construction is also critical, particularly for long-term educational use in varied settings. By addressing these factors, Echo-Teddy can bridge the gap between controlled intervention settings and real-world application, making it a scalable, adaptable, and impactful tool for special education.

\section{Conclusion}

Expert interviews on the initial prototype of Echo-Teddy identified four critical areas for enhancing its effectiveness in supporting social communication for autistic students. First, minimizing response time and ensuring stable interaction were emphasized as key factors in maintaining engagement. Second, the robot’s physical form and nonverbal behavior modules required careful refinement to align with the sensory and social preferences of autistic children. Third, experts highlighted the need for tailored interaction scenarios and reinforcement strategies that adapt to specific learning goals and contextual factors. Finally, considerations for practical distribution and long-term maintenance were noted as essential for ensuring scalability and usability in real-world educational and home settings. These findings reinforced the importance of designing Echo-Teddy to extend everyday communication experiences beyond controlled environments, making it a useful companion in inclusive classrooms, therapy sessions, and home interactions.

Future research will focus on addressing these expert recommendations by enhancing Echo-Teddy’s engineering performance in key areas such as response time, motion control, and camera-based recognition. Further development will also explore interaction scenarios specifically tailored to the social and emotional characteristics of autistic students, ensuring that conversations and behavioral prompts align with their needs. Another priority is testing compatibility with Augmentative and Alternative Communication (AAC) input methods, allowing for multi-modal interactions that accommodate a broader range of communication preferences. The research team will also investigate modular design options, offering variations such as teddy bears or other animal forms to improve user preference and increase long-term engagement.

To evaluate Echo-Teddy’s real-world impact, longitudinal studies in inclusive classrooms will assess its role as a mediator between autistic and non-autistic students, examining how it facilitates peer interaction and social skill development over time. Through these refinements, Echo-Teddy aims to become a practical and scalable tool in special education, helping children and adolescents on the autism spectrum develop meaningful communication skills in both structured and naturalistic settings.


\newpage

\section{Appendix 1: Component Specifications}
\label{appendix}

\renewcommand{\arraystretch}{1.2}
\begin{table}[hbt!]
    \centering
    \scriptsize
    \renewcommand{\arraystretch}{1.2} 
    \setlength{\extrarowheight}{1pt}  
    \begin{tabularx}{\textwidth}{XXXXX} 
        \hline
        \textbf{Category} & \textbf{Component Type} & \textbf{Model Name} & \textbf{Company Model Name} & \textbf{Specifications} \\
        \hline
        Echo-Teddy Speaker & Sound Card & - & IN-U71CW (IN NETWORK) & USB Virtual 7.1 Channel \\
        Echo-Teddy Speaker & Digital Amplifier Module & PAM8403 & SZH-AMBO-006 (SMG) & 2 x 3W \\
        Echo-Teddy Speaker & Speaker Module & - & FQ-024 (SMG) & 40mm 8$\Omega$ 2W Magnetic Speaker \\
        Microphone & Pin Microphone & - & P5HD (Joytron) & USB \\
        Push Button & Push Button & - & ZAS-BU-003 (PRC) & 12x12mm 4-Pin, SMD Type \\
        Dot Matrix & Dot Matrix Module & MAX7219 & SZH-DMBN-002 (SMG) & 8x8 LED Matrix \\
        Motor & Servo Motor & - & MG995 (SMG) & Size: 40.7 × 19.7 × 42.9mm \\
        LED & LED & - & 5BB4SC00 (DAKWANG) & 5mm LED, Red Color \\
        Single Board Computer & Raspberry Pi & - & Raspberry Pi 5 & RAM 4GB, Quad-Core CPU \\
        \hline
    \end{tabularx}
\end{table}

\end{document}